\input{aipcheck}

\documentclass[
    ,final            
  ]
  {aipproc}

\layoutstyle{6x9}

\begin{document}

\title{Mesons in the nuclear Medium}

\classification{13.60.Le, 14.40.-n, 25.20.-x}
\keywords      {omega meson, medium modification, mass}

\author{Martin Kotulla, for the CBELSA/TAPS collaboration}{
  address={II. Physikalisches Institut, Heinrich-Buff Ring 16, 35392 Giessen,
  Germany} 
}



\begin{abstract}
We discuss recent experimental results on the modification of hadron
properties in a nuclear medium. Particular emphasis is placed on an $\omega$
production experiment
performed by the CBELSA/TAPS collaboration at the ELSA accelerator. The data
shows a smaller $\omega$ meson mass together with a significant increase
of its width in the nuclear medium.
\end{abstract}

\maketitle


\section{Introduction}

Hadrons are composite systems bound by the strong interaction. At short ranges
($r \le 0.1$~fm) or high energies, the theory of strong interaction, QCD,
is very well tested and confirmed by perturbative methods.
However, at larger distances or at energies in the range of the lowest lying 
hadrons (e.g. the nucleon mass $\approx 1$~GeV), the
perturbative picture breaks down and the full complexity manifests itself in a
many body structure of gluons, valence quarks and sea quarks. 
Consequently, 
the mass of a hadron consisting of light quarks (u,d,s) can not be deduced
from the mass of the elementary valence quarks alone. The mass balance for the
proton, e.~g., is 938~MeV/c$^2$ comparing with $\approx$~12~MeV/c$^2$ for all three current
quark masses. Therefore, almost the entire mass must be generated by the
strongly 
interacting binding field or the properties of the QCD vacuum.
 Embedding a hadron
into a nucleus, another strongly interacting environment, should
necessarily affect (and alter) its mass. Moreover, chiral symmetry is another
player which is at the 
very heart of QCD. For massless quarks, which is very close to nature in case
of $u,d,s$ quarks, right
and left handed quark fields decouple. A consequence in the hadron spectrum
would be the degeneracy of opposite parity states,
 which is neither realized for
baryons nor for mesons. The reason is the spontaneous breakdown of chiral
symmetry, indicated by its measure or order parameter, the non-zero vacuum
expectation value of the $\bar{q}q$ operator (chiral condensate). 
However, model calculations
suggest a significant temperature and density dependence of the chiral
condensate pointing to a partial restauration of chiral symmetry. The
experimental observable consequence would be a change of hadron masses towards
degeneracy of opposite parity states. \\
The change of hadron properties when embedded in the nuclear medium is an
intensively 
debated field in hadron physics. Today, the qualitative and quantitative
behaviour 
of the mass of vector mesons is at hand, triggered by a few next generation
experiments \cite{trnka:mass,naruki:mass,damjanovic:mass}.
The general method to measure a modification of a hadron 
mass is to produce a sufficiently short lived meson in a nuclear
environment either via a heavy ion collision or in an elementary reaction on a
nucleus. The $\rho$ meson is a possible probe, since its $c\tau = 1.3$~fm
is smaller than the size of a nucleus (a few fm). Otherwise, the $\rho$ meson
is very broad (150~MeV), which makes its distinction from a $\pi \pi$
background difficult. The $\omega$ meson is an alternative probe, its
$c\tau = 23$~fm is small enough to have a sufficiently high probability for a
decay inside a nucleus and its vacuum width (8.5~MeV) promises a well defined
signal.  

\section{The $\omega$ meson in the nuclear medium}

\begin{figure}
  \includegraphics[width=.65\textwidth]{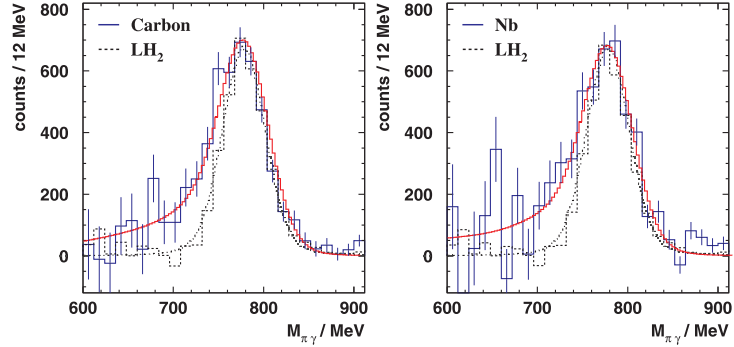}
  \caption{After background subtraction and for $p_\omega \le
    500$~MeV/c. Left: The $\omega$ line-shape (blue solid histogram) on C in 
    comparison with a LH$_2$ reference signal (dashed histogram) and a Geant
    response simulation (dashed line). A BUU transport calculation with a 16\%
    downward mass shift reproduces the data (red solid line)
    \cite{muehlich:mass}. Right: The same
    analysis for Nb.}
  \label{fig:mass}
\end{figure}

The experiment was performed at the electron stretcher accelerator (ELSA) in
Bonn using the Crystal Barrel and TAPS calorimeters (see \cite{trnka:mass} for
more details and \cite{klein:cipanp} for a facility overview). The $\omega$
mesons were produced via the reaction $\gamma A 
\rightarrow \omega X$ and detected through their $\omega \rightarrow \pi^0
\gamma$ decay. This channel offers a 1000~times higher branching ratio than
$e^+e^-$ measurements and a unique isolated signal since the $\rho$ meson
decay is suppressed by a factor 100, whereas di-lepton decays are of the same
order for the $\rho$ and $\omega$. Otherwise, a distortion of the $\pi^0
\gamma$ invariant mass by a possible re-scattering of the strongly interacting
$\pi^0$ meson in the nucleus has to be investigated very carefully and it has
been shown that the
effect in the $\omega$ mass region is negligible
\cite{johan:omega,muehlich:rescatter}. The resulting spectrum of $\pi^0
\gamma$ events shows a pronounced $\omega$ peak on a smooth background, which
is mainly originating from incompletely measured $\pi^0 \pi^0$ events when one
photon escaped detection. The background subtracted signal is shown in
fig.~\ref{fig:mass} for C and Nb targets. Details of the background
subtraction can be found in \cite{trnka:mass,trnka:phd}.  
The line-shape on the nuclear
targets shows a significant tail on the low invariant mass side when compared
to the reference shapes on LH$_2$ or a GEANT simulation. The extraction of an
in-medium signal is less straight forward, since the decay of the
$\omega$ meson occurs at different densities of the nucleus resulting in an
average density of $\approx$~60\%$\rho_0$ for Nb and a little less for the C
target. Therefore, models need to be used to separate quantitatively the
integrated mass distribution m($\rho$) and to unfold the spectral shape at
normal nuclear matter density $\rho_0$. As an
example, fig.~\ref{fig:mass} shows the expected line-shape within a BUU
calculation with an $\omega$ mass 
lowered by 16\% at $\rho_0$. Calculations with an 8\% dropping mass describe
the measured data similarily well. Although, the experimental resolution of
FWHM=55~MeV together with the theoretical models limit the extraction of a
precise in-medium signal, a clear evidence of a dropping mass scenario for
the $\omega$ meson can be concluded. \\
   
\begin{figure}
  \includegraphics[width=.65\textwidth]{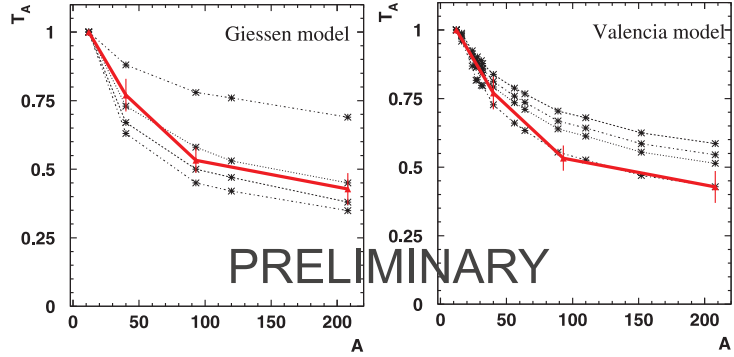}
  \caption{PRELIMINARY: The transparency ratio normalized on $^{12}$C for the
    measured data 
  (solid red line) in comparison with calculations. The best fit
  line corresponds to a width of $\Gamma(\rho_0,p=0$~MeV/c$)=40$~MeV for the
  Giessen model \cite{muehlich:width} and
  $\Gamma(\rho_0,p=750$~MeV/c$)=95$~MeV for the Valencia 
  model \cite{oset:width}. }
  \label{fig:width}
\end{figure}

The extraction of the $\omega$ width in the nuclear medium from the
measurement of the mass distribution is limited by the experimental resolution
and the unfolding of the nuclear density distribution at which the $\omega$
decays. Fortunately, the measurement of the $\omega$ absorption as a function
of the nuclear size allows a  straight forward way to extract the (inelastic)
cross 
section of the $\omega$ meson which can be related to its width
\cite{muehlich:width, oset:width}. Here, the transparency ratio $T_A$ is
introduced, which is a measure of the probability for the $\omega$ meson to
leave the nuclear without absorption. Fig.~\ref{fig:width} shows $T_A$ for the
measured data in comparison to two model calculations.   
As a preliminary result, $\Gamma_{\mbox{Gi}}(\rho_0,p=0$~MeV/c$)=40$~MeV
\cite{muehlich:width} and
$\Gamma_{\mbox{Val}}(\rho_0,p=750$~MeV/c$)=95$~MeV \cite{oset:width} can
found. The width in the Giessen calculation has been 
computed as a function of the $\omega$ momentum and agrees at higher momenta
with the Valencia value
($\Gamma_{\mbox{Gi}}(\rho_0,p=750$~MeV/c$)=105$~MeV). The intuitive 
expectation of a significant broadening of the $\omega$ in nuclear matter as a
consequence of the opening of additional partial channels (e.g. $\omega N
\rightarrow NN$) can be confirmed. This topic will be elaborated in a
forthcoming publication.





\bibliographystyle{aipproc}   

\bibliography{kotulla_cipanp.bbl}

\IfFileExists{\jobname.bbl}{}
 {\typeout{}
  \typeout{******************************************}
  \typeout{** Please run "bibtex \jobname" to optain}
  \typeout{** the bibliography and then re-run LaTeX}
  \typeout{** twice to fix the references!}
  \typeout{******************************************}
  \typeout{}
 }

\end{document}